# Frequency-domain spin-wave logic gate based on three-magnon confluence

*Vladimir D. Bessonov[1], Stephanie R. Lake[2], Bikash Das Mohapatra[2], Jannis Bensmann[3], Georg Schmidt[2,4,5], Rudolf Bratschitsch[3], Sergej O. Demokritov[1], and Vladislav E. Demidov[1]*

[1]Institute of Applied Physics, University of Münster, 48149 Münster, Germany

[2]Institut für Physik, Martin-Luther-Universität Halle-Wittenberg, 06120 Halle, Germany

[3]Institute of Physics and Center for Nanotechnology, University of Münster, 48149 Münster, Germany

[4]Interdisziplinäres Zentrum für Materialwissenschaften, Martin-Luther-Universität Halle-Wittenberg, 06120 Halle, Germany

[5]Halle-Berlin-Regensburg Cluster of Excellence CCE

ABSTRACT: Spin-wave-based information processing provides a promising platform for low-power nanoscale computing by exploiting coherent magnon dynamics and intrinsic nonlinear interactions. A key advantage of wave-based processing for efficient miniaturization is the ability to perform operations in the frequency domain without requiring separate propagation paths for the input and output signals. Here, we demonstrate a frequency-domain spin-wave logic gate based on resonant, non-degenerate three-magnon confluence in an yttrium iron garnet nanowaveguide.

Under resonant conditions, two input spin waves at different frequencies efficiently generate an output spin wave at the sum frequency, which occurs only when both input signals are present, thereby directly implementing an AND logic operation. We further show that the output intensity scales linearly with each input intensity and is therefore proportional to their product, enabling analog multiplication of two signals. These results demonstrate nonlinear magnon confluence as a versatile mechanism for frequency-domain spin-wave logic and analog signal processing.

Conventional computing architectures rely on the transport of electrical charge in semiconductor devices. The growing demand for faster and more energy-efficient information processing has driven the exploration of alternative physical carriers of information. Magnonics is an emerging field that exploits spin waves and their quanta, magnons, to transmit and manipulate information at the nanoscale[1,2]. Spin waves in low-damping, nanometer-thick magnetic insulators, such as yttrium iron garnet (YIG), offer a promising platform for low-loss information transport[3-5]. They provide multiple degrees of freedom for information encoding and manipulation[6-7], can be transmitted and controllably routed without a static bias magnetic field[8], and can be efficiently amplified through parametric processes[9]. Furthermore, strong nonlinear interactions between spin waves provide additional functionalities that are particularly attractive for implementing unconventional computing paradigms[10-14].

In recent years, a wide range of magnonic logic devices has been developed based on the interference and nonlinear interactions of spin waves[15-24]. In most of these approaches, logic operations are performed in the spatial domain: input signals are launched into different propagation channels and their interference or interaction within a magnetic structure determines the output state. These concepts have enabled the realization of basic logic functions as well as more complex magnonic computing circuits, including magnonic half-adders[19], neural networks[11], and co-processors[22]. More recently, reconfigurable magnonic logic gates have been demonstrated that can implement multiple logic functions within a single device[23]. These developments have established an important foundation for future integrated spin-wave-based electronics. However, processing information in the spatial domain also presents significant challenges. In particular, it requires precise control of spin-wave propagation paths within magnonic networks, which can complicate the design and integration of increasingly complex systems. As the number of interconnected functional blocks increases, these requirements can impose limitations on the scalability, flexibility, and reconfigurability of magnonic computing architectures.

An alternative approach that can help overcome these challenges is to perform signal-processing operations directly in the frequency domain, using distinct frequency channels to represent independent input and output signals[10,13]. Frequency-domain processing provides a complementary means of manipulating spin-wave information, in which the result of a logic operation is determined by nonlinear interactions between signals at different frequencies rather than by the interference of equal-frequency signals in real space. In particular, nonlinear magnon-magnon interactions can mediate frequency conversion allowing multiple input channels to interact and generate an output at a different frequency. Such processes therefore provide a direct physical mechanism for implementing frequency-domain logic and basic signal-processing functions.

Among the nonlinear interactions existing in magnetic systems, three-magnon confluence[25-31] is particularly attractive for this purpose. In this process, two magnons combine to generate a third magnon. Thus, when two input signals have frequencies $f_1$ and $f_2$, the confluence process generates an output signal at the sum frequency $f_1 + f_2$, with the output state determined by those of the two input magnons. Despite the evident potential of this process for implementing multi-input signal-processing operations, its experimental observation is challenging because it requires the simultaneous fulfillment of energy and momentum conservation, which is difficult to achieve for spin waves with a strongly nonlinear frequency-wavenumber dependence. Recently, it was shown that these conservation conditions can be satisfied in YIG nanowaveguides by exploiting a combination of different thickness modes resulting in the resonant second-harmonic generation[30]. However, this demonstration was limited to the frequency-degenerate case, in which the two magnons undergoing confluence have equal frequencies ($f_1 = f_2$), and is therefore not suitable for implementing multi-input operations.

In this Letter, we demonstrate the realization of non-degenerate resonant three-magnon confluence in YIG nanowaveguides over a broad range of frequency differences between the two input signals. We show that the confluence process directly implements the functionality of a logical AND operation in the frequency domain. We further demonstrate that the output intensity scales with the product of the input intensities, enabling analog multiplication in the same device. Finally, we demonstrate AND-gate operation using pulsed input signals and show that the process can be operated on the nanosecond time scale. These results establish non-degenerate three-magnon confluence as a mechanism for frequency-domain spin-wave logic and analog signal processing.

Figure 1 presents the experimental configuration used to investigate nonlinear spin-wave interactions in magnetic nanowaveguides. The waveguide has a width of 500 nm and a thickness of 80 nm. It is fabricated from a YIG film grown by pulsed-laser deposition on a <111>-oriented gadolinium gallium garnet (GGG) substrate, followed by patterning using electron-beam lithography and a lift-off process[32]. A static magnetic field, $H_0$, is applied perpendicular to the longitudinal axis of the waveguide, defining the propagation geometry of the quasi-surface spin waves. Spin waves are excited inductively using a 200-nm-thick Au antenna with a width of 500 nm. The antenna is driven simultaneously by two monochromatic microwave signals with frequencies $f_1$ and $f_2$, which serve as the two input channels of the investigated device. Through a three-magnon confluence process, the nonlinear interaction between the two excited spin waves generates a sum-frequency component at $f_1 + f_2$, which serves as the output signal of the device. Note that all input and output signals propagate in the same waveguide without the need for physically separated paths.

The input and nonlinearly generated spin waves are characterized using micro-focus Brillouin light-scattering (BLS) spectroscopy[33]. For the BLS measurements, laser light with a wavelength of 473 nm and an optical power of 0.25 mW is focused onto the waveguide surface using a microscope objective lens with 100× magnification and a numerical aperture of 0.9, producing a diffraction-limited probing spot, as illustrated in Fig. 1. The interaction between the incident light and the spin-wave excitations results in frequency-shifted scattered light. By measuring the intensity of the scattered component (BLS intensity) as a function of frequency, the spectral characteristics of the spin-wave modes can be determined. The measured BLS intensity is proportional to the intensity of the corresponding spin-wave mode, enabling the detection and characterization of spin waves generated through nonlinear interactions at different frequencies.

First, we use a single-frequency excitation regime to determine the resonance conditions for the three-magnon confluence process in the studied device[30]. A static magnetic field of $\mu_0 H_0 = 50$ mT is applied perpendicular to the longitudinal axis of the waveguide, and spin waves are excited by a monochromatic signal with a power of 0.1 mW. The excitation frequency $f_e$ is varied from 2.2 to 2.7 GHz, covering the frequency range of the fundamental spin-wave mode at this field, as discussed below. The color map in Fig. 2a shows BLS spectra acquired in the waveguide at a distance of 10 μm from the antenna axis as a function of $f_e$. The strong signal along the $f_d = f_e$ line, where $f_d$ denotes the detection frequency, corresponds to the linear excitation of spin waves by the antenna. In addition, a sharp resonance at $f_d = 2f_e$ is observed when $f_e = f_r = 2.46$ GHz, exhibiting a spectral width of less than 0.02 GHz. As shown by the spectrum in Fig. 2b, the intensity of the double-frequency signal under these conditions becomes comparable to that of the directly excited spin wave, demonstrating the high efficiency of the nonlinear process, in which two excited magnons combine into a magnon with twice the frequency. A second, weaker resonance is also observed at $f_e = 2.56$ GHz. Owing to its substantially lower intensity, this resonance is less suitable for practical applications. The physical origin of this secondary resonance is discussed elsewhere[30].

The dispersion diagram in Fig. 3a illustrates the physical mechanism underlying the observed resonance. Efficient confluence of two magnons at frequency $f$ into a magnon at frequency $2f$ requires conservation of both energy and linear momentum. Consequently, the wavenumber of the resulting magnon must be twice that of the initial magnons. Figure 3a shows the dispersion relation of the fundamental spin-wave mode, $f_F(k)$, together with that of the first-order thickness mode, $f_H(k)$, calculated using an analytical theory[34] and the approach developed in Ref.[35]. For the calculations, we use a YIG saturation magnetization of $\mu_0 M_s = 175$ mT and an exchange constant of 3.25 pJ/m. To determine whether the resonance conditions can be satisfied, we plot $2f_F(2k)$ as a function of $k$ and look for its intersections with $f_H(k)$. As seen in Fig. 3a, the two dependencies intersect at only one point, corresponding to a frequency of 4.96 GHz for the resulting magnon and, therefore, to an initial-magnon frequency of 2.48 GHz. This value is in good agreement with the experimentally observed resonance frequency $f_r = 2.46$ GHz.

The degenerate magnon confluence process described above can be useful for technical applications related to frequency multiplication. However, this process cannot be directly employed in information-processing devices that require more than one input signal, such as logic gates. To enable these applications, we propose using a non-degenerate confluence process, in which the two input magnons have different frequencies, $f_1 = f_r - \Delta f$ and $f_2 = f_r + \Delta f$ (see Fig. 3b). Conservation of both energy and linear momentum then requires the corresponding wavenumbers to satisfy $k_1 = k_r - \Delta k$ and $k_2 = k_r + \Delta k$. This condition requires the dispersion relation $f_F(k)$ to be linear in the vicinity of $f_r$, which can only be satisfied approximately.

To demonstrate that, despite the restrictions discussed above, the non-degenerate process can be implemented in practice over a substantial frequency range around $f_r$, we apply two signals to the input antenna at frequencies $f_1 = f_r - \Delta f$ and $f_2 = f_r + \Delta f$, and measure the BLS spectra while increasing $\Delta f$ from 0 to 0.2 GHz. The spectra shown in Fig. 4a clearly demonstrate the feasibility of the non-degenerate resonant process. Although the peak at $f_1 + f_2 = 2f_r$ decreases in intensity with increasing $\Delta f$, likely due to the nonlinearity of $f_F(k)$, it remains pronounced even at $\Delta f = 0.2$ GHz, as shown by the spectrum in Fig. 4b. Thus, the system under study enables a substantial frequency separation between the two input channels without significant degradation of the nonlinear conversion efficiency.

It is important to emphasize that resonant non-degenerate three-magnon confluence directly implements the functionality of a logic AND operation. Indeed, applying either input signal alone cannot initiate the confluence process because the resonance conditions are not satisfied at either $f_1$ or $f_2$. The confluence process can therefore be initiated only when both signals are applied simultaneously. This behavior is illustrated by the BLS spectra in Figs. 4c and 4d, recorded upon applying the signals at $f_1 = f_r$ - 0.2 GHz and $f_2 = f_r$ + 0.2 GHz, respectively. As seen from these data, no confluence signal is observed in the high-frequency part of the spectrum in either case.

We now examine in more detail how the intensity of the confluence peak at the sum frequency $f_1 + f_2$ depends on the intensities of the input signals at frequencies $f_1$ and $f_2$. To this end, we fix the power $P$ of one of the input signals at 0.1 mW and vary the power of the other signal from 0 to 0.15 mW. The resulting dependences shown in Figs. 5a and 5b demonstrate that the confluence-peak intensity is linearly proportional to the intensity of each input signal when the other is kept fixed, and is therefore proportional to their product. This behavior is expected for an AND operation, which is analogous to multiplication of two binary values, 0 and 1. Furthermore, the observed linear dependences indicate that the confluence process can also be used for analog multiplication of two input signals, an important function in analog signal-processing applications.

Finally, we analyze the operation of the proposed mechanism in the pulsed regime. We apply the input signals at frequencies $f_1$ and $f_2$ in the form of 500-ns-long pulses, with a relative temporal shift of 250 ns (Fig. 6a), and record the temporal profile of the output confluence signal at the sum frequency $f_1 + f_2$ (Fig. 6b). As seen from these data, a nonzero output signal is observed only during the time interval in which the input pulses overlap, as expected for a logical AND operation. Additionally, the output pulse at the sum frequency exhibits sharp edges, and its duration is only about 3 ns shorter than the overlap time of the input pulses at $f_1$ and $f_2$. This observation indicates that the finite development time of the confluence process does not significantly limit the operating speed of the proposed device.

In conclusion, we have demonstrated highly efficient non-degenerate resonant three-magnon confluence in YIG nanowaveguides over a broad range of frequency differences between the input signals. This nonlinear interaction enables a direct implementation of a logical AND operation in the frequency domain, while its intrinsic multiplicative response provides analog multiplication within the same device. Moreover, we demonstrate operation on the nanosecond time scale, highlighting the potential of this mechanism for high-speed magnonic devices. These results establish non-degenerate three-magnon confluence as a versatile platform for frequency-domain spin-wave logic and analog signal processing, opening a route toward compact multifunctional magnonic devices that combine digital and analog functionalities.

FIGURES

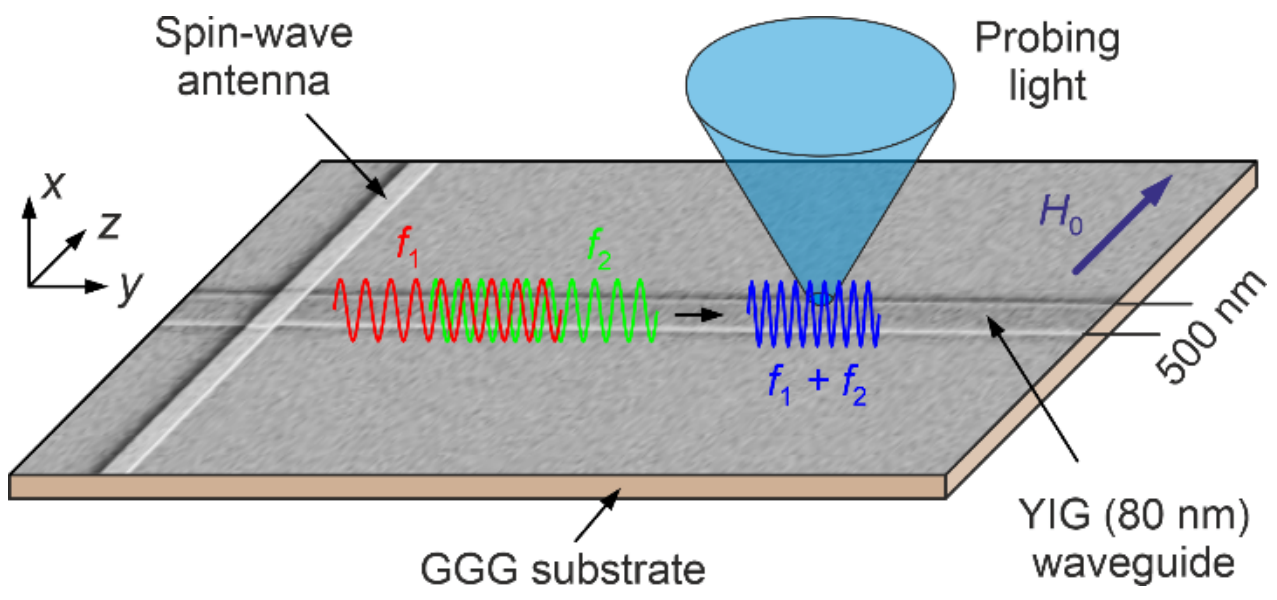

**Figure 1.** Schematic of the experiment. Micro-focus BLS is used to study the resonant generation of a spin wave at the sum frequency $f_1 + f_2$, resulting from the nonlinear interaction of two spin waves with frequencies $f_1$ and $f_2$ in an 80-nm thick YIG waveguide with a width of 500 nm.

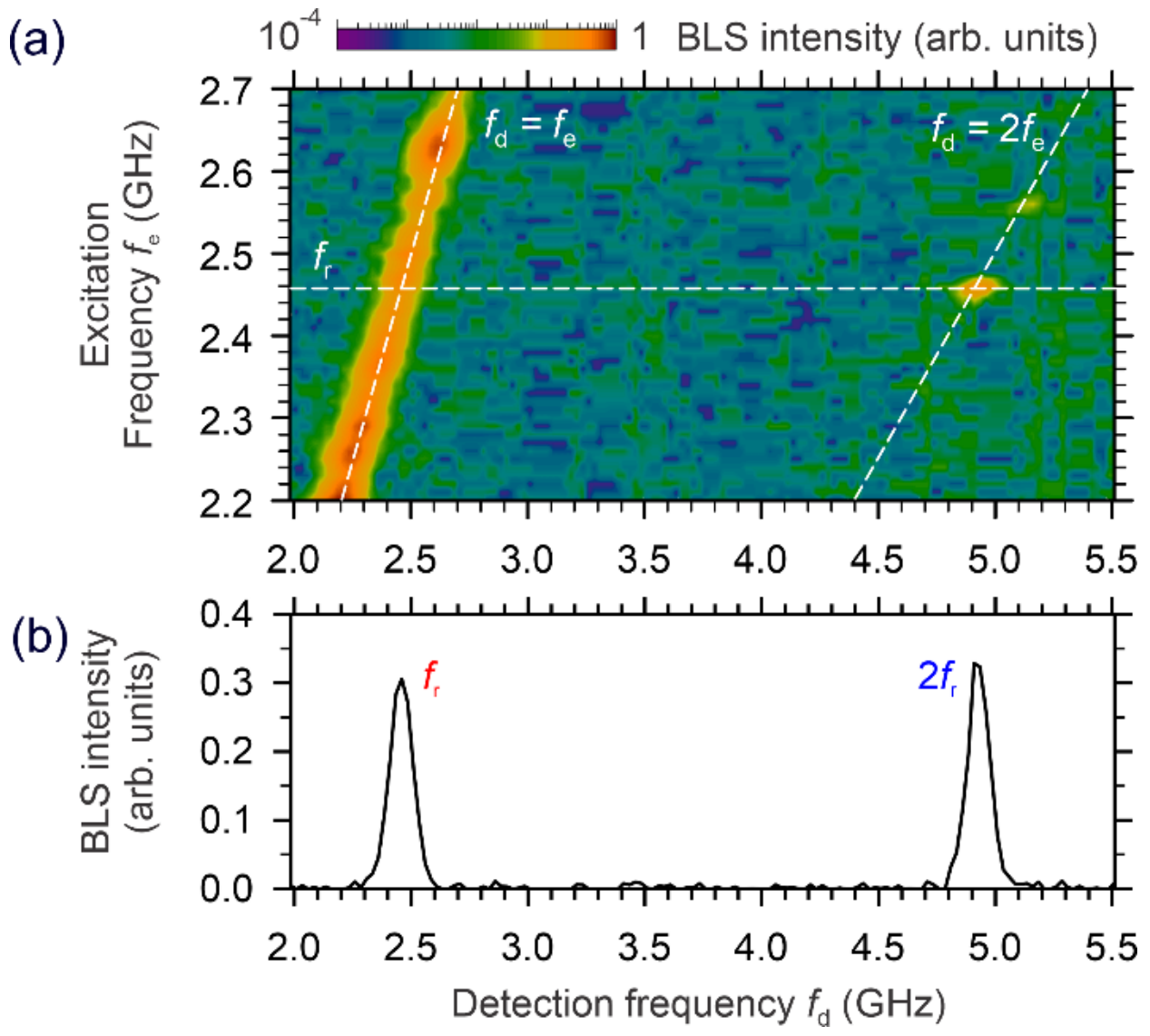

**Figure 2.** Determination of the resonance conditions for the three-magnon confluence process under single-frequency excitation. (a) BLS spectra measured in the waveguide at a distance of 10 μm from the antenna as a function of the excitation frequency $f_e$. $f_r$ denotes the resonant frequency. (b) BLS spectrum measured at $f_e = f_r$. The data are recorded under an applied magnetic field of $\mu_0 H_0 = 50$ mT.

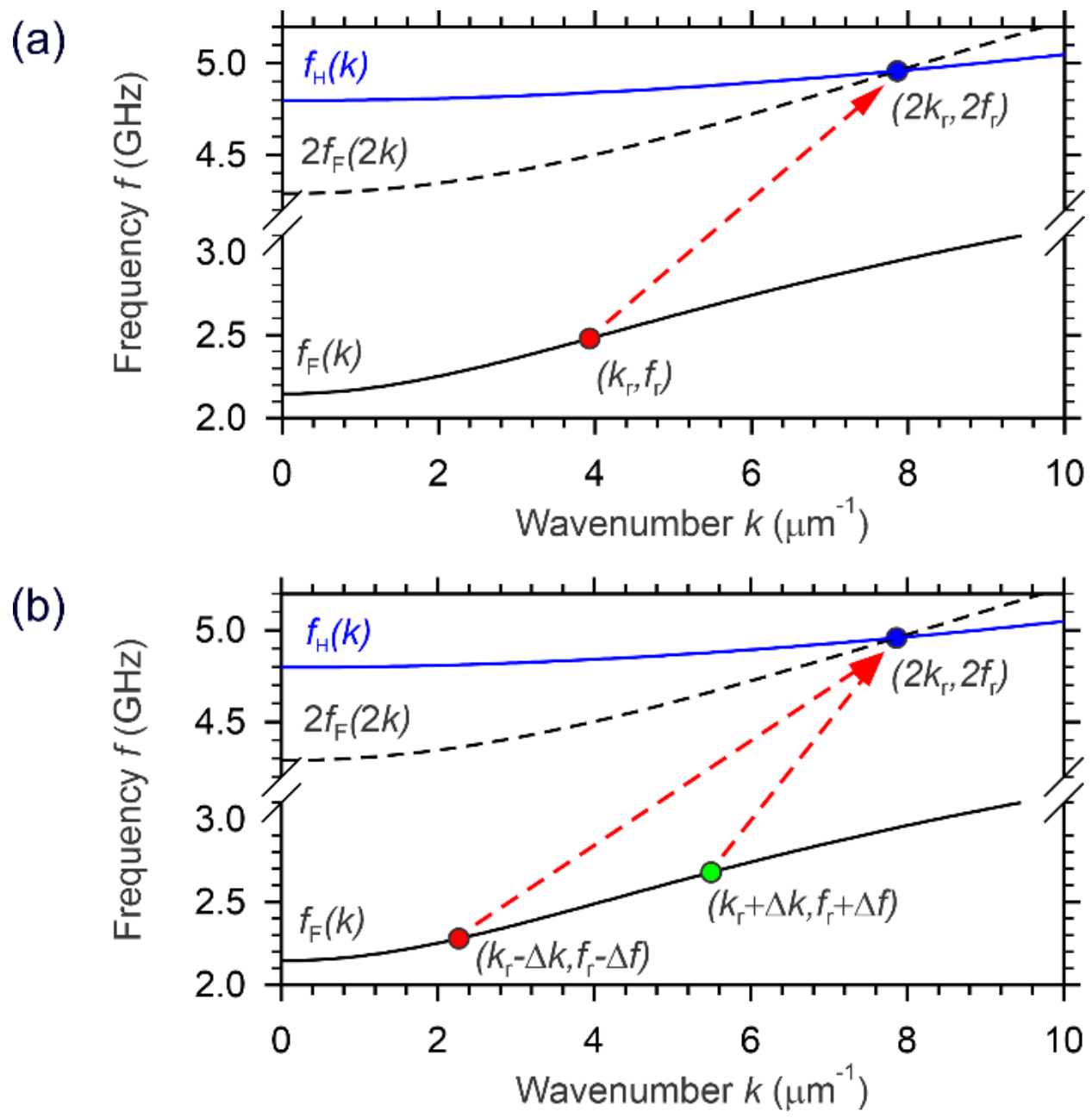

**Figure 3.** Calculated dispersion diagrams illustrating the degenerate (a) and non-degenerate (b) resonant magnon confluence process. $f_F(k)$ and $f_H(k)$ denote the dispersion relations of the fundamental and first-order thickness spin-wave modes, respectively. The intersection of $f_H(k)$ and $2f_F(2k)$ corresponds to the simultaneous fulfillment of the energy and momentum conservation conditions for the three-magnon confluence process. The calculations are performed for a static magnetic field of $\mu_0 H_0$ = 50 mT.

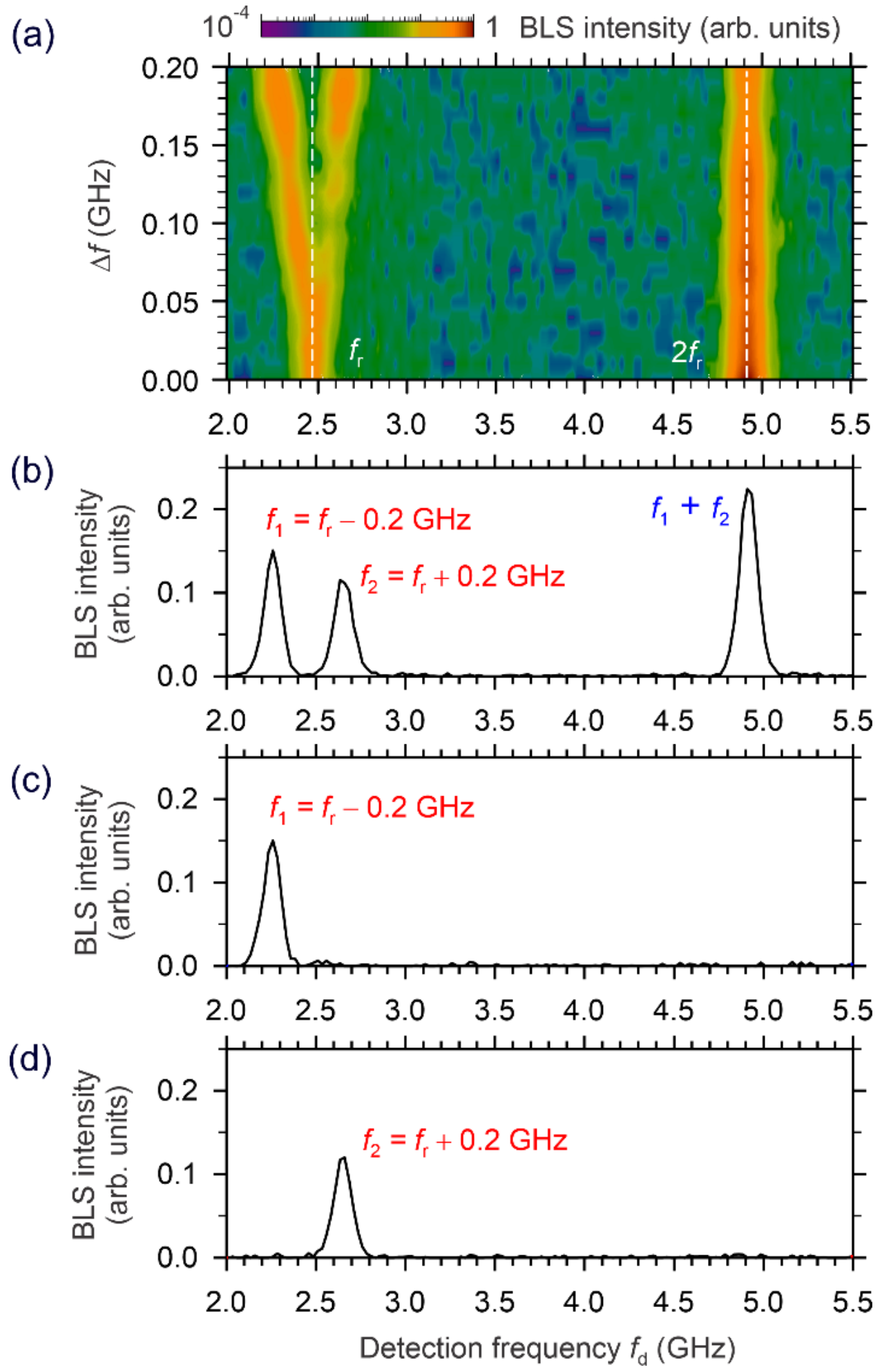

**Figure 4.** Non-degenerate three-magnon confluence process under two-frequency excitation at $f_1 = f_r - \Delta f$ and $f_2 = f_r + \Delta f$. (a) BLS spectra measured in the waveguide at a distance of 10 μm from the antenna as a function of the excitation frequency detuning $\Delta f$. (b) BLS spectrum measured at $\Delta f = 0.2$ GHz. (c) and (d) Reference BLS spectra measured under single-frequency excitation at $f_1 = f_r - 0.2$ GHz and $f_2 = f_r + 0.2$ GHz, respectively. The data are recorded under an applied magnetic field of $\mu_0 H_0 = 50$ mT.

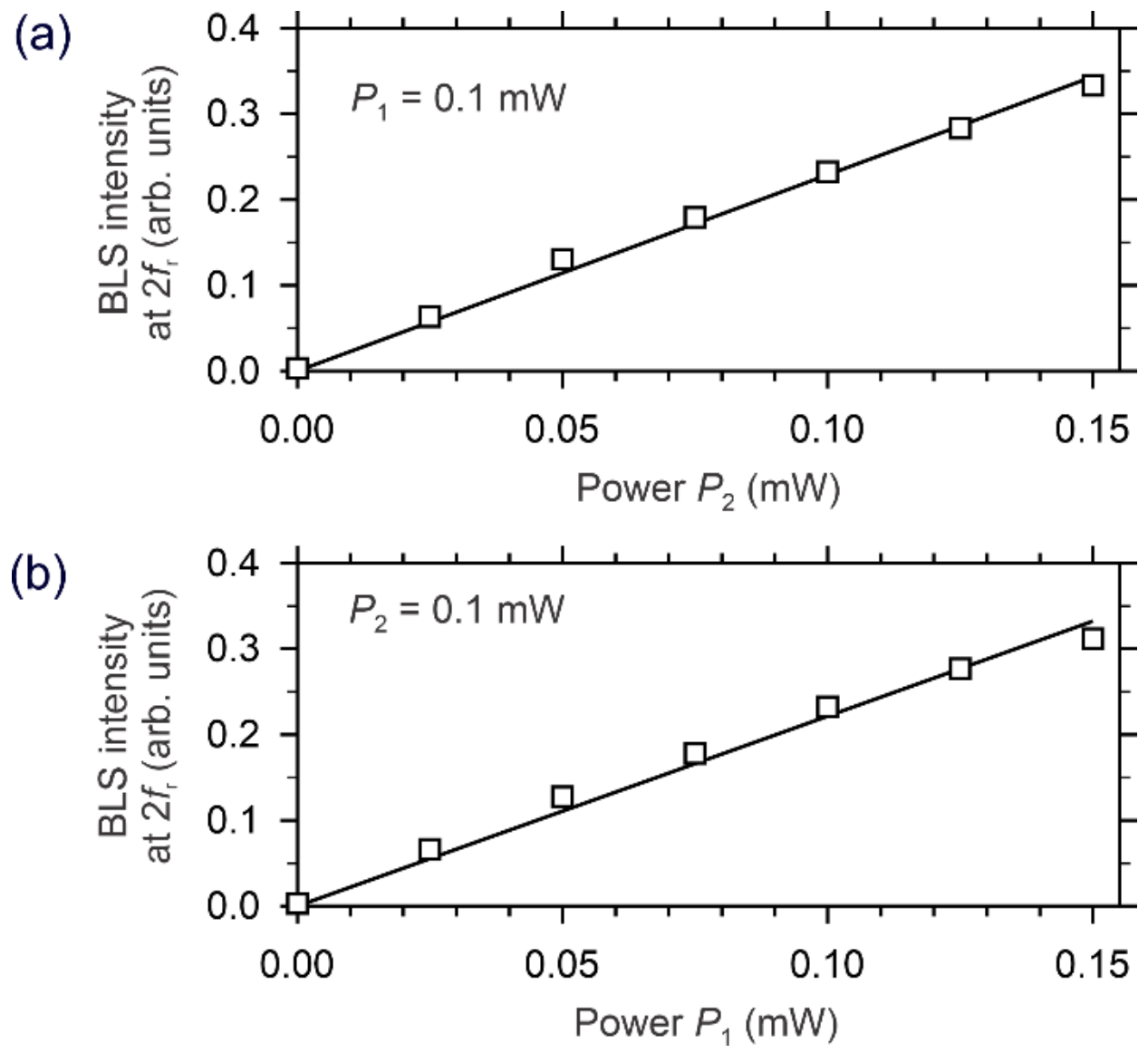

**Figure 5.** Intensity of the confluence peak at the sum frequency as a function of the excitation powers $P_1$ and $P_2$ of the waves at $f_1 = f_r$ - 0.2 GHz and $f_2 = f_r$ + 0.2 GHz, respectively. (a) $P_1$ is fixed at 0.1 mW, while $P_2$ is varied from 0 to 0.15 mW. (b) $P_2$ is fixed at 0.1 mW, while $P_1$ is varied from 0 to 0.15 mW. Symbols show the experimental data, and solid lines represent the corresponding linear fits. The data are recorded under an applied magnetic field of $\mu_0 H_0$ = 50 mT.

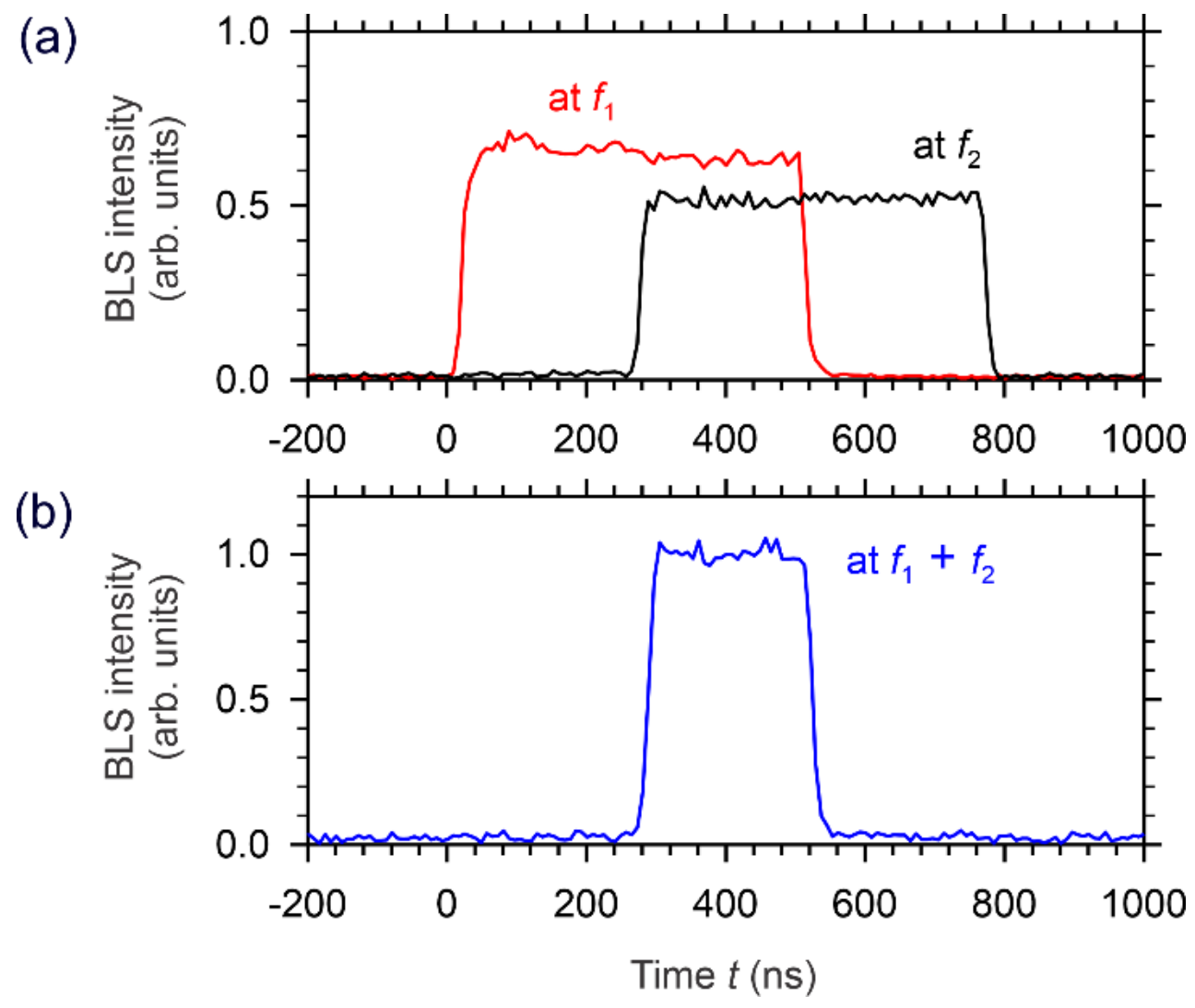

**Figure 6.** Non-degenerate confluence process measured in the pulsed regime. (a) Temporal traces of the excited spin-wave pulses measured at a distance of 10 μm from the antenna for excitation frequencies $f_1 = f_r$ - 0.2 GHz and $f_2 = f_r$ + 0.2 GHz, as labeled. (b) Temporal trace of the confluence signal at the sum frequency. The data are recorded under an applied magnetic field of $\mu_0 H_0$ = 50 mT.

AUTHOR INFORMATION

**Corresponding Author**

**Vladislav E. Demidov** – Institute of Applied Physics, University of Münster, 48149 Münster, Germany; E-Mail: demidov@uni-muenster.de

**Author Contributions**

V.D.B. performed measurements and data analysis. S.R.L and B.D.M. grew and characterized the films and performed nanofabrication. J.B. performed data analysis. G.S., R.B., S.O.D. and V.E.D.

formulated the experimental approach and supervised the project. V.E.D. wrote the original manuscript. All authors discussed the results and commented on the manuscript.

**Notes**

The authors declare no competing financial interest.

ACKNOWLEDGMENT

G.S. and V.E.D. acknowledge support by the Deutsche Forschungsgemeinschaft (DFG, German Research Foundation) – Project-ID 529812702. V.D.B, J.B., R.B. and S.O.D acknowledge support by the Deutsche Forschungsgemeinschaft (DFG, German Research Foundation) – Project-ID 433682494 – SFB 1459. G.S. acknowledges support by the Deutsche Forschungsgemeinschaft (DFG, German Research Foundation) as part of the German Excellence Strategy – EXC3112/1 – 533767171 (Center for Chiral Electronics).